\documentclass[12pt]{iopart}
\usepackage{graphicx,iopams}
\usepackage{esint}
\usepackage{graphicx}
\usepackage[crop=off]{auto-pst-pdf}
\usepackage{epsfig}
\usepackage{pstricks}
\usepackage{amssymb}
\usepackage{color}
\usepackage{psfrag}
\definecolor{dblue}{rgb}{0.1,0.1,0.44}
\definecolor{dgreen}{rgb}{0.2 ,0.54, 0.2}
\newcommand{\be}{\begin{equation}}
\newcommand{\ee}{\end{equation}}
\newcommand{\bea}{\begin{eqnarray}}
\newcommand{\eea}{\end{eqnarray}}
\def\th{\theta}
\begin{document} 
\bibliographystyle{unsrt} 
\title[Baryon masses in the  3-state PFT]{Baryon masses in the  three-state Potts field theory in a weak magnetic field}
\author{S.~B.~Rutkevich}
\address{Fakult\"at f\"ur Physik, Universit\"at Duisburg-Essen,  D-47058 Duisburg,
Germany}
\ead{sergei.rutkevich@uni-due.de}
\begin{abstract}
The $3$-state Potts field theory describes the scaling limit of the 
$3$-state Potts model on the two-dimensional lattice near its continuous
phase transition point. In the presence of thermal and magnetic field perturbations, the 
$3$-state Potts field theory in the ordered phase exhibits confinement of kinks, which allows 
both mesons and baryons. 
We calculate the masses of light baryons in this model in the weak confinement regime 
in leading order of the small magnetic field. In  leading order of perturbation theory,
the light baryons can be viewed as bound states of three quantum particles - the kinks, 
which move on a line and interact via a linear potential. We determine the masses of the lightest baryons 
by numerical solution of the associated non-relativistic one-dimensional quantum three-body problem.	
\end{abstract}
\section{Introduction}
Confinement of kink topological excitation is a quite common phenomenon in 
$2d$-Euclidean Quantum Field Theories (QFT), which are   invariant under some 
discrete symmetry group, and  
display a continuous order-disorder phase transition. 
After the Wick rotation, such models can be also viewed as relativistic   
 QFT with one space and one time dimension. If the system has $q$ degenerate vacua 
$|0_{\alpha}\rangle$, $\alpha=1,\ldots,q$, in the ordered phase due to the spontaneous breaking of the discrete symmetry, the particle sector of the theory should contain   kinks $K_{\alpha\beta}$, $\alpha,\beta=1,\ldots,q$ that interpolate between any two different vacua. The application of a uniform external field that  shifts to a lower value  the  energy of  the vacuum   
$|0_{q}\rangle$, lifts the degeneracy between the vacua. As a result, the vacuum $|0_{q}\rangle$ transforms into 
the true ground state, whereas the states $|0_{\alpha}\rangle$ with $\alpha=1,\ldots,q-1 $ become the non-stable false vacua.  This 
induces a long-range attractive interaction between  kinks, which, in turn,  leads to their confinement:  isolated kinks do not exist in the system any more and become bound into compound particles.  Particular realizations of this scenario in different two-dimensional field-theoretical
models have been attracted much attention in the  recent years 
\cite{DelMus98, FonZam2003,FZ06,Rut05,Del08,TakMus09,Mus11}. 

The interest on the problem  of kink confinement stems from several origins. First,  together with
particle decay, inelastic scattering and resonances \cite{ZZ2011,Zam2013}, nucleation in the false 
vacua etc. \cite{FonZam2003,Rut01,Mang2010,BKM10},  this phenomenon falls into the 
realm of non-integrable aspects of the quantum field theory and statistical mechanics, which cannot be described within the
framework of exactly solvable models. If one views quantum field theories 
as  points in the "space of local interactions"  \cite{Ma76,Cardy} that flow under the 
action of the renormalisation group, a generic point in this space would correspond to some
 non-integrable model. The famous example is the near-critical two-dimensional Ising model with non-zero magnetic field
 \cite{FonZam2003,Mang2010}.  Of course, any progress in understanding universal properties of such non-integrable models  is highly desirable.
 
The second motivation comes from  particle physics due to certain common features between confinement of kinks  and 
confinement of quarks in QCD. One example is provided by the Bethe-Salpeter equation 
 introduced in the Ising filed theory  by Fonseca and Zamolodchikov  \cite{FonZam2003}, which turns out to be very similar to the 
Bethe-Salpeter equation in 't Hooft's  
multicolor two-dimensional QCD \cite{Hooft74}, see  \cite{FZ06,FLZ09}. Note, that the kinks and 
their  bound states in the confinement regime are often referred to as quarks, mesons 
(two-kink bound states)
and baryons (three-kink bound states), respectively.

Finally, the kink confinement  can be realized in one-dimensional condensed-matter systems. 
It has  recently been experimentally observed \cite{Coldea10,Mor14}
in the one-dimensional Ising spin-chain ferromagnet cobalt niobate  
($\mathrm{Co}\mathrm{Nb}_2\mathrm{O}_6$). The magnetic structure of this compound can be described by the one-dimensional  quantum Ising spin-chain model, which is the paradigmatic model for the theory of quantum phase transitions \cite{Sach99}.

The simplest and most studied example of  kink confinement  is provided by the 
Ising Field Theory (IFT), i.e. the
Euclidean QFT, 
which describes the 
scaling limit of the two-dimensional lattice Ising model. At zero magnetic field $h=0$ in the ferromagnetic phase 
$T<T_c$, this model is characterized by a spontaneously broken
$\mathbf{Z}_2$ symmetry, has  $q=2$  degenerate vacua, and  two types of kinks $K_{1,2}$
and $K_{2,1}$, which can be viewed  as non-interacting fermions of 
 mass $m\sim (T_c-T)$. 
Only mesonic  bound states
are present in this model in the  confinement regime, when a magnetic field $h$ that explicitly 
breaks the $\mathbf{Z}_2$ symmetry is applied. 
As  $h\to0$, the  masses $\mu_n(h)$, $n=1,2,\ldots$, of the mesons  densely fill the interval $[2m,\infty)$. 
The evolution of the IFT meson masses $\mu_n(h)$ with increasing $h$ has been 
studied in great  details both numerically and analytically  \cite{McCoy78,FonZam2003,FZ06,Rut05,Rut09}.  
Two asymptotic expansions
have been  obtained for the IFT meson masses 
in the weak confinement regime $h\to 0$. 
The {\it semiclassical expansion} \cite{FZ06,Rut05} in integer powers of $h$ 
holds for the masses $\mu_n(h)$ of highly exited mesons, with $n\gg1$.
The {\it low energy expansion} 
\cite{McCoy78,FonZam2003,FZ06} in  fractional powers
of $h$ 
describes the masses of mesons with not very large values of $n$, such that $\mu_n(h)-2m\ll m$. Note, that the leading term of the 
low energy expansion can be obtained in a very simple manner from the McCoy-Wu scenario, in which 
 the meson is interpreted as a bound state of  two non-relativistic quantum
particle moving on a line and attracting one another with a linear potential.                     

The next,  reacher and  more complicated  example of the model exhibiting kink confinement is given by  the $q$-state Potts field theory.  
Being defined in two space-time dimensions only for $q\le4$, it  
represents the scaling limit of the two-dimensional $q$-state lattice Potts model \cite{Bax}  near its  continuous phase transition point. 
At zero magnetic field $h=0$, the model is invariant under  the group $S_q$ describing permutations of  $q$ colors 
and has $q$ degenerate vacua in the ordered phase. The PFT reduces to the IFT at $q=2$. 
The kinks  $K_{\alpha\beta}$, $\alpha,\beta=1,\ldots,q$, are  massive particles that interact with each other 
at short distances already when  $h=0$, if $q>2$. 
Interaction of kinks in the $h=0$ PFT  is described by a factorizable scattering matrix, as found by Chim and Zamolodchikov \cite{CZ92}. 
Application of the magnetic field in the ordered phase leads to kink confinement in the PFT, which has been studied in several  
papers \cite{Del08,LTD,Rut09P}. The symmetry analysis of the kink bound states, 
and the qualitative analysis of the evolution of their mass spectrum  were performed by 
Delfino and Grinza  \cite{Del08} . These authors have shown, that besides the mesons, also the baryonic (three-kink) bound states 
are present at $q=3,4$, and the 'tetra-quark' (four-kink) bound states appear at $q=4$ in the $q$-state PFT in the confinement regime.
Lepori, T\'oth, and Delfino \cite{LTD} have  numerically  studied the evolution of the particle mass spectrum in the 3-state PFT under variation of the
temperature and magnetic field by means of the 
Truncated Conformal Space Approach (TCSA)  \cite{YuZam91,Tak14}.
Analytic perturbative analysis of the meson mass spectrum in the $q$-state PFT has been done in paper \cite{Rut09P}, 
in which the leading terms of the both low-energy and semiclassical asymptotical small-$h$ expansions for the meson masses
were obtained.  

The purpose of the present paper is to calculate the masses of several lightest baryons in the 3-state PFT in the weak confinement regime at 
$h\to+0$. 
Following the McCoy-Wu scenario, we treat the light baryons at $h\to +0$ as bound states of three non-relativistic quantum particles (the kinks),
which move on a line and interact with each other with the linear attractive potential. This allows us to relate the  masses of light baryons 
 with  discrete energy levels
of a certain  one-dimensional quantum three-body problem,  which we approximately solve by numerical means.

The paper is organized as follows. Section \ref{Pm} contains a brief introduction into the  Potts field theory and some of its 
well-known properties 
at zero magnetic field. The weak confinement regime in the ferromagnetic 3-state PFT is discussed in Section \ref{WC}, where the 
non-relativistic quantum three-body problem,  which determines the masses of light baryons is also presented. In Section  \ref{FR}, we reduce 
this quantum three-body problem to the Fredholm integral equation, which numerical solution is described in Section \ref{NM}.
Concluding remarks are given in Section \ref{CON}.
\section{Potts field theory \label{Pm}}
The Potts field theory describes the scaling limit of the two-dimensional $q$-state Potts model
with the number of colors $q$ lying in the interval $q\le4$. 
The PFT Euclidean action can be written as \cite{CZ92,Del08,LTD}
\begin{equation} 
{\cal A}= {\cal A}_{CFT}^{(q)} -\tau\int d^2x\,\varepsilon(x)-
h\int d^2x\,\sigma_q(x)\,\,,
\label{scaling}
\end{equation}
Here ${\cal A}_{CFT}^{(q)}$ corresponds to the Conformal Field Theory, 
which is associated with the  critical point, $\varepsilon(x)$ is the 
energy density, $\sigma_\alpha(x)$ are  the spin densities, which are subject to the
constrain $\sum_{\alpha=1}^q \sigma_\alpha(x)=0$.
The couplings $\tau$ and $h$ are proportional
to the deviation of the temperature and magnetic field 
from their critical point values. 
The magnetic field $h$ is chosen to act on the $q$-th color only. At the critical point, the energy density and the spin density have the scaling dimensions $X_\epsilon^{(q)}$, $X_\sigma^{(q)}$, respectively. 

At $h=0$, the Potts model
is invariant under the permutation group $S_q$; 
at $h\ne0$ the symmetry reduces to the group $S_{q-1}$ of permutations of the colors $\alpha=1,\ldots,q-1$. 
The PFT is integrable at  $h=0$.
At $h=0$ and $\tau<0$,
the $S_q$ symmetry is  spontaneously broken: the model has $q$ degenerate 
vacua $|0_\alpha\rangle$, $\alpha=1,2,\ldots,q$, which are distinguished 
by the values of the order parameter 
$\langle \sigma_\gamma\rangle_\alpha  \equiv \langle 0_\alpha|\sigma_\gamma(x)|0_\alpha\rangle$.
The symmetry group $S_q$ acts by permutations of these vacua.

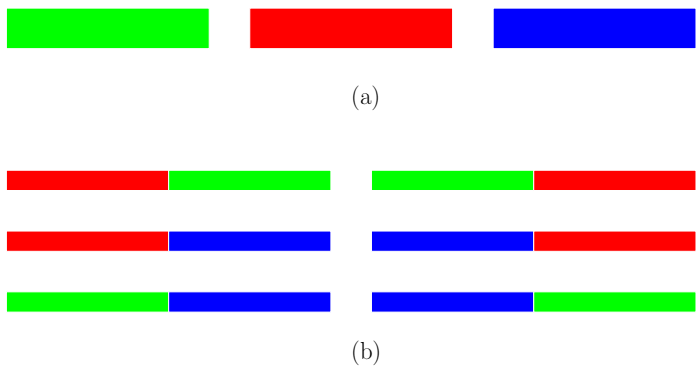
\begin{figure}[h]
\centering{
\resizebox{0.8\textwidth}{!}{
\begin{pspicture}(23,11)
\psframe[linewidth=0pt,linecolor=white,fillstyle=solid,
fillcolor=green](3,8)(8,9)
\psframe[linewidth=0pt,linecolor=white,fillstyle=solid,
fillcolor=red](9,8)(14,9)
\psframe[linewidth=0pt,linecolor=white,fillstyle=solid,
fillcolor=blue](15,8)(20,9)
\psframe[linewidth=0pt,linecolor=white,fillstyle=solid,
fillcolor=red](3,5)(7,4.5)
\psframe[linewidth=0pt,linecolor=white,fillstyle=solid,
fillcolor=green](7,4.5)(11,5)
\psframe[linewidth=0pt,linecolor=white,fillstyle=solid,
fillcolor=green](12,5)(16,4.5)
\psframe[linewidth=0pt,linecolor=white,fillstyle=solid,
fillcolor=red](16,5)(20,4.5)

\psframe[linewidth=0pt,linecolor=white,fillstyle=solid,
fillcolor=red](3,3)(7,3.5)
\psframe[linewidth=0pt,linecolor=white,fillstyle=solid,
fillcolor=blue](7,3)(11,3.5)
\psframe[linewidth=0pt,linecolor=white,fillstyle=solid,
fillcolor=blue](12,3)(16,3.5)
\psframe[linewidth=0pt,linecolor=white,fillstyle=solid,
fillcolor=red](16,3)(20,3.5)
\rput[l]{00}(11.5,0.5){{\Large{(b)}}}
\rput[l]{00}(11.5,6.8){\Large{(a)}}
\psframe[linewidth=0pt,linecolor=white,fillstyle=solid,
fillcolor=green](3,2)(7,1.5)
\psframe[linewidth=0pt,linecolor=white,fillstyle=solid,
fillcolor=blue](7,2)(11,1.5)
\psframe[linewidth=0pt,linecolor=white,fillstyle=solid,
fillcolor=blue](12,2)(16,1.5)
\psframe[linewidth=0pt,linecolor=white,fillstyle=solid,
fillcolor=green](16,2)(20,1.5)
\end{pspicture}
}}
\caption{(a) Three degenerate vacua  $|0_\alpha\rangle$, and (b) six kinks $K_{\alpha\beta}$,  $\alpha,\beta=1,2,3$  of the  $3$-state PFT at   $\tau<0$ and $h=0$.} \label{fig:figureone}
\end{figure}

In this paper, we  shall concentrate on the PFT with $q=3$ in the ordered phase  $\tau<0$.
At $h=0$ and $\tau<0$, the 3-state PFT has three degenerate vacua  $|0_\alpha\rangle$, $\alpha=1,2,3$
 shown schematically in Figure \ref{fig:figureone}a.  Its particle sector contains six kinks  $K_{\alpha \beta}(\theta)$, $\alpha,\beta=1,2,3$, which interpolate between two different vacua $\alpha$ and $\beta$, see Figure \ref{fig:figureone}b. Each  kink $K_{\alpha \beta}(\theta)$ is  a
relativistic particle having   mass $m$ and rapidity  $\theta$.  The latter  parametrizes the kink's energy $E=m\cosh \theta$ and momentum 
 $p=m \sinh \theta$. The kink mass is related with  the parameter $\tau$ as 
 $|\tau|\sim  m^{2-X_\epsilon^{(q)}}$.

In contrast to the IFT, the kinks in the 3-state PFT interact 
with each other at distances $\lesssim m^{-1}$ already at $h=0$. This interaction leads to the mutual kink scattering, which is described by the factorizable   scattering matrix. The corresponding two-kink scattering matrix  in the $q$-state PFT is 
described by the  commutation relations found by 
Chim and Zamolodchikov
 \cite{CZ92}:
\bea \label{ScM}
 K_{\alpha\gamma}(\theta_1)K_{\gamma\beta}(\theta_2)=S_0(\theta_{12})
\sum_{\delta\neq\gamma}K_{\alpha\delta}(\theta_2)K_{\delta\beta}(\theta_1)\\\nonumber +
S_1(\theta_{12})
K_{\alpha\gamma}(\theta_2)K_{\gamma\beta}(\th_1)\,,\hspace{.5cm}\alpha\neq\beta,
\label{KSc}\\
 K_{\alpha\gamma}(\theta_1)K_{\gamma\alpha}(\theta_2)=S_2(\theta_{12})
\sum_{\delta\neq\gamma}K_{\alpha\delta}(\theta_2)K_{\delta\alpha}(\theta_1)\\\nonumber+
S_3(\theta_{12})K_{\alpha\gamma}(\theta_2)K_{\gamma\alpha}(\theta_1), \label{KS}
\eea
where $\theta_{1,2}=\theta_{1}-\theta_{2}$, and
\bea
&& S_0(\theta)=\frac{\sinh\lambda\theta\,\sinh\lambda(\theta-i\pi)}
{\sinh\lambda\left(\theta-\frac{2\pi i}{3}\right)\,\sinh\lambda\left(\theta-
\frac{i\pi}{3}\right)}\,\Pi\left(\frac{\lambda\theta}{i\pi}\right),
\label{s0}\\
&& S_1(\theta)=\frac{\sin\frac{2\pi\lambda}{3}\,\sinh\lambda(\theta-i\pi)}
{\sin\frac{\pi\lambda}{3}\,\sinh\lambda\left(\theta-\frac{2 i\pi}{3}\right)}\,
\Pi\left(\frac{\lambda\theta}{i\pi}\right), \label{S1}\\
&& S_2(\theta)=\frac{\sin\frac{2\pi\lambda}{3}\,\sinh\lambda\theta}
{\sin\frac{\pi\lambda}{3}\,\sinh\lambda\left(\theta-\frac{i\pi}{3}\right)}\,
\Pi\left(\frac{\lambda\theta}{i\pi}\right),\label{s2}\\
&& S_3(\theta)=\frac{\sin\lambda\pi}{\sin\frac{\pi\lambda}{3}}\,
\Pi\left(\frac{\lambda\theta}{i\pi}\right).\label{s3}
\eea
The parameter $\lambda$ is related to $q$ via
\begin{equation}
\sqrt{q}=2\sin\frac{\pi\lambda}{3}\,,
\label{qlambda}
\end{equation}
and
\bea
&&\Pi\left(\frac{\lambda\theta}{i\pi}\right)=
\frac{\sinh\lambda\left(\theta+i\frac{\pi}{3}\right)}{\sinh\lambda(\theta-
i\pi)}\,e^{{ A}(\theta)},\\
&& { A}(\theta)=\int_0^\infty\frac{dx}{x}\,
\frac{\sinh\frac{x}{2}\left(1-\frac{1}{\lambda}\right)-
\sinh\frac{x}{2}\left(\frac{1}{\lambda}-\frac{5}{3}\right)}
{\sinh\frac{x}{2\lambda}\cosh\frac{x}{2}}\,\sinh\frac{x\theta}{i\pi}. \label{Aex}
\eea
Due to (\ref{qlambda}),  the parameter $\lambda$ takes the value unity   in the  $3$-state PFT, 
\begin{equation}
\lambda=1, \quad {\rm at} \; q=3.
\end{equation}
Note, that at $q=3$, the scaling dimensions of the energy density and the spin density operators take the values \cite{Del08}
\[
X_\epsilon^{(3)}=\frac{4}{5}, \quad X_\sigma^{(3)}=\frac{2}{15}.
\]
\section{Weak confinement in the $3$-state PFT \label{WC}}
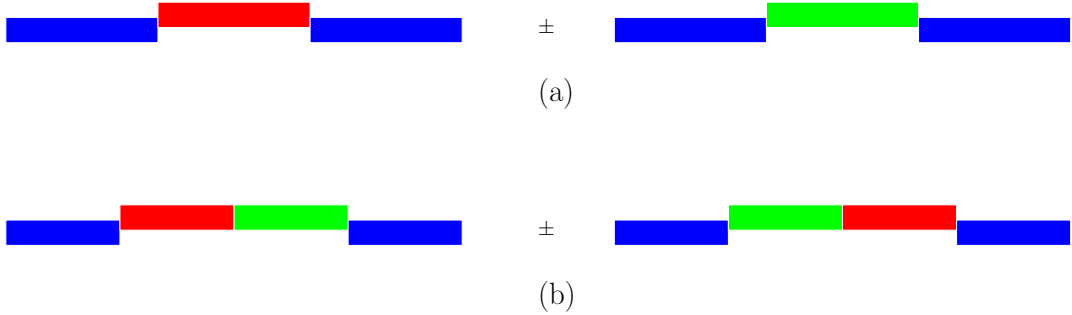
\begin{figure}[h]
\centering{
\resizebox{1\textwidth}{!}{
\begin{pspicture}(23,8)
\psframe[linewidth=0pt,linecolor=white,fillstyle=solid,
fillcolor=blue](1,5.5)(4,5)
\psframe[linewidth=0pt,linecolor=white,fillstyle=solid,
fillcolor=red](4,5.3)(7,5.8)
\psframe[linewidth=0pt,linecolor=white,fillstyle=solid,
fillcolor=blue](7,5)(10,5.5)
\psframe[linewidth=0pt,linecolor=white,fillstyle=solid,
fillcolor=blue](13,5)(16,5.5)
\psframe[linewidth=0pt,linecolor=white,fillstyle=solid,
fillcolor=green](16,5.3)(19,5.8)
\psframe[linewidth=0pt,linecolor=white,fillstyle=solid,
fillcolor=blue](19.,5)(22,5.5)
\rput[l]{00}(11.5,4.){\Large{(a)}}
\psframe[linewidth=0pt,linecolor=white,fillstyle=solid,
fillcolor=blue](1,1.5)(3.25,1)
\psframe[linewidth=0pt,linecolor=white,fillstyle=solid,
fillcolor=red](3.25,1.3)(5.5,1.8)
\psframe[linewidth=0pt,linecolor=white,fillstyle=solid,
fillcolor=green](5.5,1.3)(7.75,1.8)
\psframe[linewidth=0pt,linecolor=white,fillstyle=solid,
fillcolor=blue](7.75,1)(10,1.5)
\psframe[linewidth=0pt,linecolor=white,fillstyle=solid,
fillcolor=blue](13,1)(15.25,1.5)
\psframe[linewidth=0pt,linecolor=white,fillstyle=solid,
fillcolor=green](15.25,1.3)(17.5,1.8)
\psframe[linewidth=0pt,linecolor=white,fillstyle=solid,
fillcolor=red](17.5,1.3)(19.75,1.8)
\psframe[linewidth=0pt,linecolor=white,fillstyle=solid,
fillcolor=blue](19.75,1)(22,1.5)
\rput[l]{00}(11.5,0){{\Large{(b)}}}
\rput[l]{00}(11.5,1.3){${\pm}$}
\rput[l]{00}(11.5,5.3){$\displaystyle\pm$}
\end{pspicture}
}}
\caption{(a) Meson,  and (b) baryon bound states of kinks in the   $3$-state PFT at   $\tau<0$ and 
a small $h>0$.} \label{fig:figure3}
\end{figure}

Application of a weak magnetic field $h>0$ along the third direction decreases the energy of the 
vacuum $|0_3\rangle$, lifting the  degeneracy between the  
vacuum $|0_3\rangle$ and the remaining ones $|0_{1,2}\rangle$. 
The vacuum $|0_3 \rangle$ becomes the true ground state of the system, and 
the states $|0_{1,2} \rangle$ become  false (metastable) vacua. 
In the leading order in $h$, the energy density difference between the ground state $|0_3 \rangle$
and the metastable vacua $|0_{1,2} \rangle$ reads  \cite{Del08}
\begin{eqnarray}
&&\Delta \mathcal{E}=\mathcal{E}_{\alpha}-\mathcal{E}_{3}= f_0+O(h^2),
\quad \alpha=1,2, \\
&&f_0=\frac{3 b h}{2},\label{strt}
\end{eqnarray}
with some positive  $b$.
This energy shift leads to a long-range linear attractive potential between kinks, which 
causes their confinement  into mesonic 
and baryonic bound states \cite{Del08} .
There are two series of mesonic states $\pi_n^{(\kappa)}$,
$$\pi_{n}^{(\kappa)}=K_{31}K_{13}+(-1)^\kappa K_{32}K_{23},$$
and two series of baryonic  states $p_n^{(\kappa)}$,
$$p_{n}^{(\kappa)}=K_{31}K_{12}K_{23}+(-1)^\kappa K_{32}K_{21}K_{13},$$
 which differ by their parity $\kappa=0,1$, see Figure \ref{fig:figure3}.

It was shown in \cite{Rut09P} that 
  the masses of the lightest mesons 
 $\mu_n^{(\kappa)}$, $n=1,2,\ldots$, at small $h>0$ can be determined to
 leading order in $h$ from the Schr\"odingier eigenvalue problem 
 \begin{equation*}
\left[ -\frac{\partial_{x_1}^2}{2m} -\frac{\partial_{x_2}^2}{2m}+2m-\mu_n^{(\kappa)}+(x_2-x_1)f_0\right]\psi_n^{(\kappa)}(x_1,x_2)=0,
 \end{equation*}
 where  $-\infty<x_1<x_2<\infty$, and $f_0$ is the string tension given by (\ref{strt}).
  The eigenfunction   $\psi_n^{(\kappa)}(x_1,x_2)$ must be 
 translationally invariant, i.e.,    
\begin{equation}\label{tra}
 \psi_n^{(\kappa)}(x_1,x_2)=\psi_n^{(\kappa)}(x_1+X,x_2+X)\quad {\textrm {for all  }}X,
\end{equation}
vanish fast enough at $(x_2-x_1)\to+\infty$, and satisfy the boundary conditions 
 \begin{eqnarray*}
 && \psi_n^{(1)}(x_1,x_2)\big|_{x_1=x_2}=0, \\
 &&  ( \partial_{x_1}-\partial_{x_2})\psi_n^{(0)}(x_1,x_2)\big|_{x_1=x_2}=0.
  \end{eqnarray*}
  It was shown in \cite{Rut09P}, that
 these boundary conditions follow  from the small-$\theta$ asymptotics of the two-kink scattering 
  matrix (\ref{ScM})-(\ref{s3}) at $q=3$. The translation invariance requirement (\ref{tra}) guaranties, that 
  the wave function  $\psi_n^{(\kappa)}(x_1,x_2)$ describes  the meson at rest. 
  
  The resulting mass spectrum of the light mesons reads  \cite{Rut09P}:
  \begin{eqnarray*}
  \mu_n^{(1)}=2m +f_0^{2/3} \,m^{-1/3}\,z_n+O(f_0^{4/3}),\\
    \mu_n^{(0)}=2m +f_0^{2/3} \,m^{-1/3}\,z'_n+O(f_0^{4/3}),
   \end{eqnarray*} 
   where $(-z_n)$ and $(-z'_n)$ are the zeroes of the Airy function ${\rm Ai}(x)$
   and its derivative ${\rm Ai}'(x)$, respectively.
   
  In order to determine  the masses of  the lightest baryons $M_n^{(\kappa)}$ at small $h>0$, 
  we shall apply a similar strategy. The subsequent analysis  is essentially  based on the 
 experience gained from  the  perturbative study \cite{FZ06,Rut09} of the meson mass spectrum in the Ising field theory. 
 
    In  leading order in $h\to+0$, the baryon can by treated as
  a bound state of three kinks, whereas   the $N$-kink fluctuations with $N>3$ should contribute to the 
  baryon state at higher orders in $h$.    In the three-kink approximation, the state 
  $|B^{(\kappa)}(P)\rangle$ of a baryon 
 having the momentum $P$ and parity $\kappa$ can  be described in the position-space 
 representation by the wave function 
 \begin{eqnarray}\label{Psi}
 \Psi^{(\kappa)}(x_1,x_2,x_3;P)=\langle K_{31}(x_1)K_{12}(x_2)K_{23}(x_3)|B^{(\kappa)}(P)\rangle=\\
(-1)^\kappa \langle K_{32}(x_1)K_{21}(x_2)K_{13}(x_3)|B^{(\kappa)}(P)\rangle,\nonumber
 \end{eqnarray}
 where $x_\alpha$, $\alpha=1,2,3$, are the spatial 
 coordinates of the kinks, 
  \begin{equation}\label{domx}
 \fl \{x_1,x_2,x_3\}\in \Gamma\subset \mathbb{R}^3, \quad \Gamma=\{x_1,x_2,x_3\in \Gamma|-\infty<x_1<x_2<x_3<\infty\}.
 \end{equation}  
 The wave function (\ref{Psi}) should vanish fast enough as $(x_3-x_1)\to\infty$, and obey the 
 following symmetry properties 
 \begin{eqnarray}\label{sym}
 \Psi^{(\kappa)}(x_1+X,x_2+X,x_3+X;P)= e^{iPX}\,\Psi_P^{(\kappa)}(x_1,x_2,x_3;P),\\
  \Psi^{(\kappa)}(-x_3,-x_2,-x_1;-P)=(-1)^\kappa \,\Psi^{(\kappa)}(x_1,x_2,x_3;P),\nonumber
  \end{eqnarray}
for arbitrary $X$. 
  
  Since the masses $M_n^{(\kappa)}$ of baryons with small enough $n=1,2\ldots$ at $h\to+0$  only slightly exceed 
 $ 3m$, $M_n^{(\kappa)}-3m\sim h^{2/3}$, the kinetic energies of three kinks forming a baryon 
 with zero momentum $P=0$ should be 
 small compared with their mass $m$. This allows one to treat such three kinks  as 
  non-relativistic quantum particles.
 These  particles interact  with each other with the linear potential 
    \begin{equation}\label{V3}
  V(x_1,x_2,x_3)=(x_3-x_1)\,f_0
  \end{equation}
  at large distances $(x_3-x_1)\gg m^{-1}$.
The above potential  does not depend on the 
coordinate $x_2$ of the middle kink, which separates domains of the metastable phases 1 and 2  
characterized by  the same energy density $\mathcal{E}_1=\mathcal{E}_2$, see Figure \ref{fig:figure3}b.

Accordingly, the masses $M_n^{(\kappa)}$ of the lightest baryons  $n=1,2,\ldots$ can be 
   determined in the leading order in $h\to+0$ from the three-particle 
   Schr\"odinger partial differential equation
    \begin{equation}\label{Sch}
\left[ -\sum_{j=1}^3\frac{\partial_{x_j}^2}{2m} +3m-M_n^{(\kappa)}+(x_3-x_1)f_0\right]\Psi_n^{(\kappa)}(x_1,x_2,x_3;0)=0,
  \end{equation}
in the domain (\ref{domx}).  
The boundary conditions for the wave function 
in the above equation at the planes $x_1=x_2$ and $x_2=x_3$ can be determined from the
small-$\theta$ asymptotes of the scattering matrix, since the rapidities of kinks forming a light 
resting baryon are small.  To this end, 
we need only the first commutation relation (\ref{ScM}),  which reduces at  $\alpha=q=3$ to the form
   \begin{eqnarray}
 K_{3\gamma}(\theta_1)K_{\gamma\beta}(\theta_2)=
S_1(\theta_{1}-\theta_{2})
K_{3\gamma}(\theta_2)K_{\gamma\beta}(\th_1),\\
     S_1(\theta)=-\exp\left[-2i
\int_0^\infty\frac{dx}{x}\,
\frac{
\sinh(x/3)}
{\sinh x}\,\sin(x\theta/\pi)
     \right],\label{S13}
  \end{eqnarray}
with  $\gamma\neq\beta$, and $\gamma,\beta\neq3$. At small rapidity $\theta$ we get
   \begin{eqnarray}\label{ff}
&&  S_1(\theta)=-1+O(\theta),\label{S13a}\\
&& K_{3\gamma}(\theta_1)K_{\gamma\beta}(\theta_2)=
-
K_{3\gamma}(\theta_2)K_{\gamma\beta}(\th_1)+O(\theta_1-\theta_2).\nonumber
  \end{eqnarray}  
The 'free-fermionic' scattering matrix element (\ref{ff}), 
 should induce the fermionic boundary conditions 
   \begin{equation}\label{bc}
 \Psi_n^{(\kappa)}(x_1,x_2,x_3;0)\big|_{x_1=x_2}=0,\quad \Psi_n^{(\kappa)}(x_1,x_2,x_3;0)\big|_{x_2=x_3}=0
  \end{equation}
for the baryon wave function in (\ref{Sch}),  
as it was explained in \cite{Rut09P} for the case of the meson wave 
functions.

For the baryon at rest, the symmetry properties (\ref{sym})  reduce to the form
 \begin{eqnarray}\label{sym0}
 \Psi^{(\kappa)}(x_1+X,x_2+X,x_3+X;0)= \Psi_P^{(\kappa)}(x_1,x_2,x_3;0),\\
 \Psi^{(\kappa)}(-x_3,-x_2,-x_1;0)=(-1)^\kappa \,\Psi^{(\kappa)}(x_1,x_2,x_3;0).\nonumber
  \end{eqnarray}
\section{Reducing of the eigenvalue problem  to the Fredholm integral equation \label{FR}}
Let us proceed  to the rescaled variables:
\begin{eqnarray}
&&x_j=( m f_0)^{-1/3} \xi_j, \quad j=1,2,3,\\
&&M_n^{(\kappa)}=3m+{f_0^{2/3}}{m^{-1/3}}\,\epsilon_n^{(\kappa)},\label{Mdim}\\
&& \Psi_n^{(\kappa)}(x_1,x_2,x_3;0) =( m f_0)^{1/3}\, \Phi_n^{(\kappa)}(\xi_1,\xi_2,\xi_3). 
\end{eqnarray}
Then, equation (\ref{Sch}) takes the form
    \begin{equation}\label{Sch1}
\left( -\frac{1}{2}\sum_{j=1}^3{\partial_{\xi_j}^2} -\epsilon_n^{(\kappa)} +\xi_3-\xi_1\right)\Phi_n^{(\kappa)}(\xi_1,\xi_2,\xi_3)=0,
  \end{equation}
for  $-\infty<\xi_1<\xi_2<\xi_3<\infty$.
We shall use  the following Ansatz for the solution of this equation:
\begin{equation}  \label{Ans}
\fl \Phi_n^{(\kappa)}(\xi_1,\xi_2,\xi_3)=\int_0^\infty d p\, g^{(\kappa)}(p)\,
f^{(\kappa)}\left[p\left(\xi_2-\frac{\xi_1+\xi_3}{2}\right)\right] \,
{\rm {Ai}}\left(\xi_3-\xi_1+\frac{3p^2}{4 }-\epsilon_n^{(\kappa)}\right),
\end{equation}
where $f^{(0)}(z)=\cos z $,   $f^{(1)}(z)=\sin z $, ${\rm {Ai}}(\xi)$ is the Airy function, 
and $g^{(\kappa)}(p)$ is some unknown function.

One can easily check, that the right-hand side of (\ref{Ans}) with an arbitrary 
function $g^{(\kappa)}(p)$ providing convergence of the integral in $p$ gives a 
 solution 
of  equation (\ref{Sch1}).   
The function $\Phi_n^{(\kappa)}(\xi_1,\xi_2,\xi_3)$ defined by (\ref{Ans}) also satisfies  the equalities
 \begin{eqnarray}\label{symA}
 \Phi^{(\kappa)}(\xi_1+\xi,\xi_2+\xi,\xi_3+\xi)= \Phi^{(\kappa)}(\xi_1,\xi_2,\xi_3), \quad \xi\in \mathbb{R},\\
 \Phi^{(\kappa)}(-\xi_3,-\xi_2,-\xi_1)=(-1)^\kappa \,\Phi^{(\kappa)}(\xi_1,\xi_2,\xi_3),\nonumber
  \end{eqnarray}
which guarantee (\ref{sym0}). 

We require, further, that $\Phi^{(\kappa)}(\xi_1,\xi_2,\xi_3)$ vanishes fast enough 
as $(\xi_3-\xi_1)\to+\infty$ in the domain  $-\infty<\xi_1<\xi_2<\xi_3<\infty$.
To satisfy also the boundary condition (\ref{bc}), one ought to require that the
 function $g^{(\kappa)}(p)$  is the solution of the  linear homogeneous integral equation
 \begin{equation}\label{Fr}
 \int_0^\infty dp\, U^{(\kappa)}(\xi,p;\epsilon_n^{(\kappa)})\,g^{(\kappa)}(p)=0
 \end{equation}
on the half-line $0<\xi<\infty$, with the asymmetric kernel
 \begin{equation}\label{ker}
U^{(\kappa)}(\xi,p;\epsilon)= f^{(\kappa)}\left({p \,\xi}/{2}\right) \,{\rm {Ai}}\left(\xi+\frac{3p^2}{4 }-\epsilon\right).
 \end{equation}
After the change of variables 
\begin{eqnarray}
&&\xi=\frac{u}{1-u}, \quad p=\left[\frac{4 v}{3(1-v)}\right]^{1/2},\\
&&g^{(\kappa)}(p) ={\phi^{(\kappa)}(v)}\,{(3v)^{1/2}\,(1-v)^{3/2}},\nonumber
\end{eqnarray}
the integral equation (\ref{Fr}) takes the form
 \begin{equation}\label{Fr1}
 \int_0^1 dv\, K^{(\kappa)}\left(u,v;\epsilon_n^{(\kappa)}\right)\,\phi^{(\kappa)}(v)=0,
 \end{equation}
for $0\le u\le 1$, whose kernel
 \begin{eqnarray}
\fl
K^{(\kappa)}(u,v;\epsilon)= 
f^{(\kappa)}\left(\frac{u\, v^{1/2} }{(1-u)[3(1-v)]^{1/2}}\right)\,
{\rm {Ai}}\left(\frac{u}{1-u}+
\frac{v}{1-v}-\epsilon\right)\label{ker1}
 \end{eqnarray}
is continuous in the square $0\le u,v\le 1$. 

Consider the Fredholm operator eigenvalue problem \cite{RS81}:
\begin{equation}\label{Fr2}
 \int_0^1 dv\, K^{(\kappa)}\left(u,v;\epsilon \right)\,g_\nu^{(\kappa)}(v;\epsilon)
 =\lambda_\nu^{(\kappa)}(\epsilon ) g_\nu^{(\kappa)}(u,\epsilon).
\end{equation}
For arbitrary complex $\epsilon$,  the spectrum $\lambda_\nu^{(\kappa)}(\epsilon ) $, $\nu=1,2,\ldots,\infty$ is discrete
with the limiting point $\lambda_\nu^{(\kappa)}(\epsilon )\to 0$ at  $\nu\to \infty$.
The baryon masses $M_n^{(\kappa)}$, $n=1,2,\ldots$ are determined through (\ref{Mdim}) by the  solutions of  the equation
\begin{equation}\label{la}
\lambda_{\nu(n)}^{(\kappa)}(\epsilon)\big|_{\epsilon=\epsilon_n^{(\kappa)}}=0,
\end{equation}
with some $\nu(n)$.
\section{Numerical solution of equation (\ref{la})\label{NM}}
For the numerical solution, the Fredholm integral operator in equation (\ref{Fr2}) was 
replaced by the  $\epsilon$-dependent square $N\times N$-matrix 
$\mathcal{K}_{ij}^{(\kappa)}(\epsilon,N)$,
\begin{eqnarray}\label{KM}
\mathcal{K}_{ij}^{(\kappa)}(\epsilon,N)=\frac{K^{(\kappa)}\left(u_i,v_j;\epsilon \right)}{N},\\
u_i=\frac{i}{N-1}, \quad v_j=\frac{j}{N-1},\quad i,j=0,1,2,\ldots,N-1.\nonumber
\end{eqnarray}
 In the limit $N\to\infty$, eigenvalues $\Lambda_\nu^{(\kappa)}(\epsilon,N)$
 of this matrix 
should approach  the eigenvalues  of the 
Fredholm integral equation (\ref{Fr2}),
\begin{equation}\label{laN}
\lim_{N\to\infty}\Lambda_\nu^{(\kappa)}(\epsilon,N)=\lambda_\nu^{(\kappa)}(\epsilon).
\end{equation}
\begin{table}[b]
\begin{tabular}{c|cccc}
\hline
$\nu$ & 1 & 2 &3  &4  \\
 \hline
 $\Lambda_\nu^{(0)}(0,31)$ &  0.104289 & -0.00416232 & 0.000225285 & -0.0000140359 \\
$\Lambda_\nu^{(0)}(0,101)$ & 0.102381& -0.00407641& 0.000220799& -0.0000137643\\
$\Lambda_\nu^{(1)}(0,31)$ & 0.0104795& -0.000685642& 0.0000447833& $ -3.02246\times10^{-6}$  \\
$\Lambda_\nu^{(1)}(0,101)$ & 0.0107229& -0.000701738& 0.0000458517& $-3.09608\times10^{-6}$\\
  \hline
  \end{tabular}
  \caption{Four initial eigenvalues  $\Lambda_\nu^{(\kappa)}(0,N)$ of the $N\times N$-matrices $\mathcal{K}_{ij}^{(\kappa)}(\epsilon,N)$ 
  at $\epsilon=0$ for $N=31,101$ and $\kappa=0,1$.}
\label{tab1}
\end{table}

\begin{figure}[h]
\centering{
\resizebox{\textwidth}{!}{
\begin{pspicture}(24,30)
\psline[linewidth=1pt, linecolor=blue]{-}(14,28)(16,28)
\psline[linewidth=1pt, linecolor=red]{-}(14,26.5)(16,26.5)
\psline[linewidth=1pt, linecolor=green]{-}(14,25)(16,25)
\rput[l]{00}(16.5,26.6){${\rm {Re}}\,\Lambda_\nu^{(0)}(\epsilon,101)$}
\rput[l]{00}(16.5,28.05){${\rm {Re}}\,\Lambda_\nu^{(0)}(\epsilon,31)$}
\rput[l]{00}(16.5,25.1){${\rm {Im}}\,\Lambda_\nu^{(0)}(\epsilon,31)$}
\rput[cm](11.5,22){%
\includegraphics
[width=
1\linewidth]
{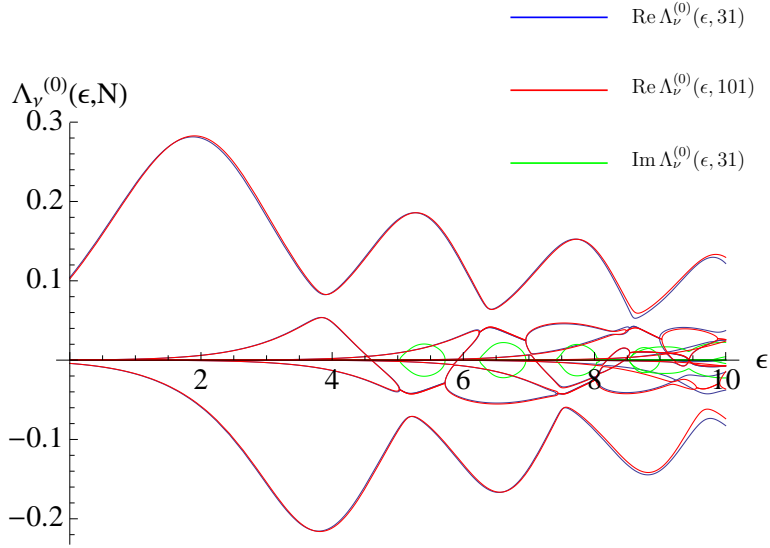}%
}
\psline[linewidth=1pt, linecolor=blue]{-}(14,14)(16,14)
\psline[linewidth=1pt, linecolor=red]{-}(14,12.5)(16,12.5)
\psline[linewidth=1pt, linecolor=green]{-}(14,11)(16,11)
\rput[l]{00}(16.5,14.1){${\rm {Re}}\,\Lambda_\nu^{(1)}(\epsilon,31)$}
\rput[l]{00}(16.5,12.6){${\rm {Re}}\,\Lambda_\nu^{(1)}(\epsilon,101)$}
\rput[l]{00}(16.5,11.1){${\rm {Im}}\,\Lambda_\nu^{(1)}(\epsilon,31)$}
\rput[cm](11.5,7){%
\includegraphics
[width=
1\linewidth]
{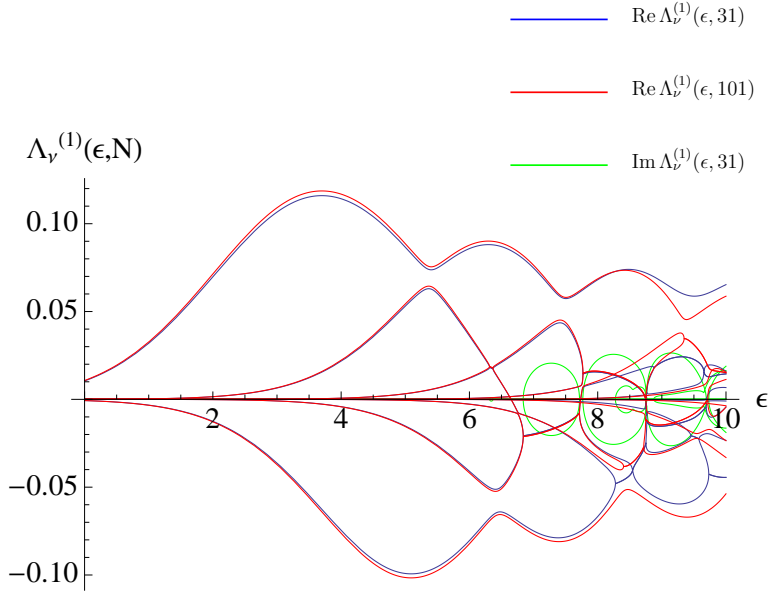}%
}
\rput[l]{00}(12.5,0.8){{\Large{(b)}}}
\rput[l]{00}(12.5,16){{\Large{(a)}}}
\end{pspicture}
}}
\caption{Evolution of the eigenvalues of the discretized versions of the
Fredholm  integral equation (\ref{Fr2})
with increasing $\epsilon$  with matrix dimensions $N=31$, and $N=101$  
for the baryon parities (a) $\kappa=0$, and (b) $\kappa=1$.
\label{spe}} 
\end{figure}

The matrix $\mathcal{K}_{ij}^{(\kappa)}(\epsilon,N)$ 
was diagonalized numerically for given $N$. The evolution of its eigenvalues
upon increasing $\epsilon$ 
is plotted in Figure \ref{spe} for the matrix dimensions $N=31$ and $N=101$ . For complex eigenvalues, 
both real and imaginary parts are displayed. 

The obtained numerical results indicate a very fast decrease of the  absolute values  of the eigenvalue 
$\Lambda_\nu^{(\kappa)}(\epsilon,N)$
with increasing $\nu$, at fixed $\epsilon$ and $N$. 
For $\epsilon=0$, this is illustrated in Table~\ref{tab1}.  Accordingly, only two eigenvalues 
corresponding to $\nu=1,2$ can be distinguished from zero in Figure \ref{spe} at $\epsilon=0$.
As soon as $\epsilon$ increases, subsequent eigenvalues with $\nu=3,4,\ldots$ become larger in  
absolute values, deviating  one by one from the abscisses in Figures \ref{spe}a,b.
 
Another important feature of the spectral patterns shown in Figure \ref{spe} is that the blue
 and red  curves, which correspond to the matrix dimensions $N=31$, and $N=101$,
respectively,  remain rather
close to each other at small enough $\epsilon\lesssim 8$. Due to 
(\ref{laN}), one should expect, that these patterns are also close to the 
asymptotical $N\to\infty$ spectra,
i.e. to the spectra 
$\lambda_{\nu(n)}^{(\kappa)}(\epsilon)$ of the Fredholm integral equation (\ref{Fr2}). 

However, the eigenvalue $\Lambda_\nu^{(\kappa)}(\epsilon,N)$ approaches
 its limiting values $\lambda_\nu^{(\kappa)}(\epsilon)$ for $N\to \infty$  not uniformly 
in $\epsilon$: at larger $\epsilon$  one has to proceed to larger matrix dimensions $N$ to reach the 
asymptotic $N\to\infty$ value. The reason is that the large positive $\epsilon$ shifts the argument 
$z=\left(\frac{u}{1-u}+\frac{v}{1-v}-\epsilon\right)$
 of the Airy function ${\rm Ai}(z)$ in the integral kernel (\ref{ker1}) to negative $z$. Since ${\rm Ai}(z)$
oscillates at negative $z$, the kernel $K^{(\kappa)}(u,v;\epsilon)$ becomes strongly oscillating  
in the square $0<u,v<1$ at large  $\epsilon$. Therefore, as $\epsilon$ increases, 
one should  proceed to larger matrix dimensions $N$ in order to provide 'sufficient 
resolution' to approximate
 the highly oscillating kernel $K^{(\kappa)}(u,v;\epsilon)$ in the square $0<u,v<1$
 by its $N\times N$-discrete-lattice counterpart
$\mathcal{K}_{ij}^{(\kappa)}(\epsilon,N)$.

The deviations between the  blue ($N=31$)  and red ($N=101$) 
curves  in Figures  \ref{spe}a,b  become considerable  at
$\epsilon \simeq 9$ and increase  further  at larger $\epsilon$.   
This indicates that at such large 
$\epsilon\gtrsim 9$ one has to increase the matrix dimension $N$ to higher values to
be able to describe the $N\to\infty $ asymptotic spectrum 
$\lambda_\nu^{(\kappa)}(\epsilon)$.
 
Several crossings of the $\epsilon$-axis by the spectral curves are clearly seen in 
Figures  \ref{spe}a,b. 
The  first three  ones are located at 
\begin{equation}\label{cr0}
\epsilon_1^{(0)}=4.602, \quad \epsilon_2^{(0)}=5.912, \quad \epsilon_3^{(0)}=7.098, 
\end{equation} 
for  $\kappa=0$ in Figure  \ref{spe}a, and
 at 
\begin{equation}\label{cr1}
\epsilon_1^{(1)}=6.650, \quad \epsilon_2^{(1)}=7.734, \quad \epsilon_3^{(1)}=8.753, 
\end{equation} 
 for $\kappa=1$ in Figure  \ref{spe}b. Up to the digits specified in (\ref{cr0}) and (\ref{cr1}), the locations of these
 crossing points for  red  and blue  curves coincide. 
 Accordingly, we  finally get from (\ref{Mdim}) the masses of the  lightest baryons at $h\to +0$:
\begin{equation}\label{Mn_fin}
\fl M_n^{(\kappa)}=3m+{f_0^{2/3}}{m^{-1/3}}\,\epsilon_n^{(\kappa)}+O(f_0^{4/3}), \quad{\rm with } \quad \kappa=0,1, \quad n=1,2,3,\ldots,
\end{equation}
where $\epsilon_n^{(\kappa)}$ for $\nu=1,2,3$ are given by equations (\ref{cr0}), (\ref{cr1}).

It should be noted that before
 crossing the $\epsilon$-axis, the spectral curve $\lambda_\nu^{(\kappa)}(\epsilon)$ crosses 
infinitely many other spectral curves $\lambda_{\nu'}^{(\kappa)}(\epsilon)$ with large enough 
$\nu'$, since 
\[
\lim_{\nu'\to\infty} \lambda_{\nu'}^{(\kappa)}(\epsilon)=0.
\]
In fact, such crossings between two spectral curves $\lambda_{\nu}^{(\kappa)}(\epsilon)$ and
 $\lambda_{\nu'}^{(\kappa)}(\epsilon)$ are rather avoided crossings, or turning points 
 of the spectral curves. 
 \begin{figure}[h]
\centering{
\resizebox{\textwidth}{!}{
\begin{pspicture}(24,16)
\psline[linewidth=1pt, linecolor=blue]{-}(14,14)(16,14)
\psline[linewidth=1pt, linecolor=red]{-}(14,12.5)(16,12.5)
\psline[linewidth=1pt, linecolor=green]{-}(14,11)(16,11)
\rput[l]{00}(16.5,14.1){\Large${\rm {Re}}\,\Lambda_\nu^{(0)}(\epsilon,31)$}
\rput[l]{00}(16.5,12.6){\Large${\rm {Re}}\,\Lambda_\nu^{(0)}(\epsilon,101)$}
\rput[l]{00}(16.5,11.1){\Large${\rm {Im}}\,\Lambda_\nu^{(0)}(\epsilon,101)$}
\rput[l]{00}(3.4,13.4){\Large $\Lambda_\nu^{(0)}(\epsilon,N)$}
\rput[l]{00}(21,7){\Large $\epsilon$}
\rput[cm](11.5,7){%
\includegraphics
[width=
1.2\linewidth]
{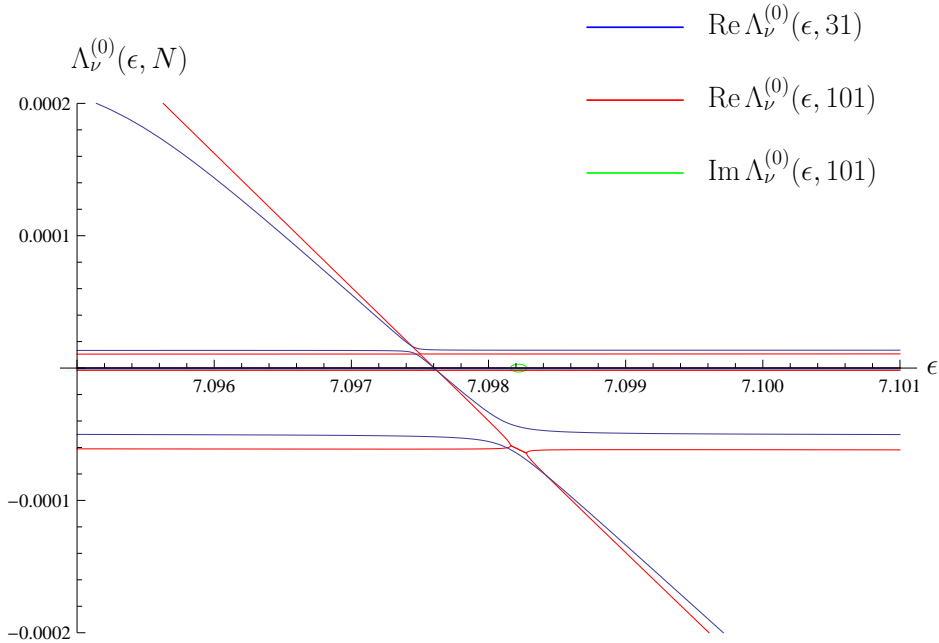}%
}
\end{pspicture}
}}
\caption{ The fine structure of the spectral curves  $\Lambda_{\nu}^{(0)}(\epsilon,N)$ near the point 
$\epsilon\approx \epsilon_3^{(0)}$. The blue curves corresponding to the matrix dimension 
$N=31$ display avoided 
crossing near the point $\epsilon=7.0982$, which transforms into two turning points in red curves 
relating  to the 
 case   $N=101$.  Between these two turning points, corresponding two eigenvalues 
  become complex and mutually conjugate.
\label{cro}} 
\end{figure}
All these features can be seen in Figure \ref{cro}, which displays the evolution of the spectral
curves  $\Lambda_{\nu}^{(0)}(\epsilon,N)$ for $N=31,101$ in the small region 
near the third crossing point $\epsilon\approx \epsilon_3^{(0)}=7.098...$
\section{Conclusions \label{CON}}
In this paper we have studied the  baryonic excitations  in  the 3-state Potts field theory
 in the weak confinement regime, which is realized in the ordered phase
in the presence of a weak magnetic field $h$  acting on the third color.  Two series of  baryons with parities $\kappa=0,1$ are allowed,  if  the
applied  magnetic field is positive, i.e. if the resulting ground state $|0_3\rangle$ becomes non-degenerate
in energy. 

To determine the masses of the light baryons  in the regime of small magnetic field $h$, 
we have applied the procedure that  was originally developed to calculate the meson masses in the IFT \cite{McCoy78,FonZam2003},
and has been 
used later \cite{Rut09P} for meson mass calculations in the  PFT. In this approach, the baryon (or the meson)   is interpreted
as a bound state of three (or two) kinks, which are treated as non-relativistic interacting quantum particles.  
The light  baryon mass spectrum  was determined from the  
stationary Schr\"odinger equation, 
which describes relative one-dimensional motion of three kinks 
with coordinates $-\infty<x_1<x_2<x_3<\infty$  interacting via the linear  confining 
potential $(x_3-x_1) f_0$.  Here $f_0>0$ is the string tension, which is proportional to the applied magnetic field $h$. 
The boundary condition for the Schr\"odingier equation at $x_1=x_2$ and $x_2=x_3$ were 
gained from the
low-momentum asymptotics of the exactly known two-kink scattering matrix  \cite{CZ92}  at $h=0$.  
The resulting mass spectrum of the  light baryons in the 
leading order in $h\to+0$ is given by equation (\ref{Mn_fin}). Several coefficients $\epsilon_n^{(\kappa)}$, with $\kappa=1,2$, and 
$n=1,2,3$ in this equation were obtained numerically, see equations  (\ref{cr0}), (\ref{cr1}).

Note, that the employed procedure is  well justified in the case of the meson mass 
spectrum in the IFT, since it reproduces the leading order of the low-energy expansion 
(see equation (5.16) in reference \cite{FZ06}), 
which was derived in the systematic perturbative approach based on the Bethe-Salpeter 
equation \cite{FonZam2003,FZ06}.

As it was shown in \cite{LTD}, the numerical TCSA method provides an alternative possibility
to study the baryon mass spectra in the 3-state PFT.  It would be interesting to perform systematic TCSA
calculations of the baryon masses in this model
at small magnetic fields, and to compare the
TCSA baryon spectra with  our results. 
\ack
I am thankful to H.~W.~Diehl for interesting discussions and numerous suggestions leading to improvement of the text.

This work was supported by Deutsche
Forschungsgemeinschaft (DFG) via Grant \newline Ru 1506/1. 

\end{document}